\documentclass[doublecol]{epl2}
\title{Competition of multiband superconducting and magnetic order in ErNi$_{2}$B$_{2}$C observed
by Andreev reflection.}

\shorttitle{Competition of multiband superconducting and magnetic order in ErNi$_{2}$B$_{2}$C}

\author{ N. L. \ Bobrov \inst{1} \and V. N. \ Chernobay \inst{1} \and Yu. G.\ Naidyuk \inst{1} \and L. V. Tyutrina \inst{1} \and  D. G. Naugle \inst{2} \and K. D. D. Rathnayaka \inst{2} \and S. L. Bud`ko \inst{3} \and P. C. Canfield \inst{3} \and I. K.\ Yanson \inst{1}}

\shortauthor{N. L. \ Bobrov \etal}

\institute{
  \inst{1} B.\ Verkin Institute for Low Temperature Physics and Engineering, National Academy of Sciences of Ukraine, 47 Lenin Ave., 61103, Kharkiv,Ukraine\\
  \inst{2} Department of Physics, Texas A\&M University, College Station TX 77843-4242,USA\\
\inst{3}Ames Laboratory, Department of Physics and Astronomy, Iowa State University, Ames, Iowa 50011, USA  }

\pacs{74.45.+c}{Proximity effects; Andreev effect; SN and SNS junctions}
\pacs{74.50.+r}{Tunneling phenomena; point contacts, weak links, Josephson effects}
\pacs{74.70.Dd}{Ternary, quaternary and multinary compounds (including Chevrel phases, borocarbides, etc.)}

\abstract{
Point contacts (PC) Andreev reflection $dV/dI$ spectra for the antiferromagnetic (T$_{\mbox{\tiny
N}}\simeq$6\,K) superconductor (T$_{c}\simeq$11\,K) ErNi$_{2}$B$_{2}$C have been measured for the two
main crystallographic directions. Observed retention of the Andreev reflection minima in $dV/dI$ up to T$_{c}$
directly points to unusual superconducting order parameter (OP) vanishing at T$_{c}$. Temperature dependence of
OP was obtained from $dV/dI$ using recent theory of Andreev reflection including pair-breaking effect. For the
first time existence of a two superconducting OPs in ErNi$_{2}$B$_{2}$C is shown. A distinct decrease of both OPs as temperature is lowered below T$_{\mbox{\tiny N}}$ is observed.  }

\begin{document}

\maketitle

\section{Introduction}
The family of quaternary nickel borocarbides superconductors $R$Ni$_2$B$_2$C, where $R$ is a rare-earth element
or Y, has attracted worldwide attention both because of a relatively high critical temperature T$_{c}$, up to
16\,K for $R$=Lu, and especially from the point of view of competition between superconducting and magnetic
ordered states in the case of $R$=Tm, Er, Ho, Dy, where energy scales for the antiferromagnetic and
superconducting  order can be varied over a wide range (see, e.\,g.,\  Refs.\,\cite{Muller,Budko06} and further
Refs.\ therein). The compound with $R$=Er and $T_c\simeq$11\,K is interesting for two reasons \cite{Muller}:
below (T$_{\mbox{\tiny N}}\simeq$6\,K) incommensurate antiferromagnetic order with spin density wave occurs and
weak ferromagnetism develops below T$_{\mbox{\tiny WFM}}\simeq$2\,K \cite{Canf96}. Both phenomena are, in
general, antagonistic to superconductivity, so that competition between superconducting  and the magnetic state
should take place in this compound. Additionally, the superconducting ground state in borocarbide
superconductors is expected to have a multiband nature \cite{Shulga0,Drechsler} with a complex Fermi surface and
different contributions to the superconducting  state by different Fermi surface sheets. Therefore, determining
the influence of these magnetic states on a possible multiband superconducting ground state or multiband order
parameter (OP) in ErNi$_2$B$_2$C is a challenge.

Previous tunneling (STM/STS) and point contact (PC) spectroscopy results have left some open questions regarding
the coexistence of superconductivity and magnetism in ErNi$_2$B$_2$C. STM/STS measurements of \cite{Wata} show a
small feature, namely, decreasing  of the superconducting gap below T$_{\mbox{\tiny N}}$ nearly within error
bars, which was not reproduced in subsequent experiments \cite{Crespo}. Early PC data on polycrystalline samples
\cite{Yanson} indicated that the superconducting gap has roughly a BCS dependence with only a shallow dip around
T$_{\mbox{\tiny N}}\simeq$6\,K. Very recent laser-photoemission spectroscopy data \cite{Baba} show the SC gap decrease (with remarkably large error bars) below the Neel temperature, but, at present the laser-photoemission spectroscopy has not enough resolution to go deeper in details.

In this paper we report our detailed directional PC Andreev reflection measurements on single crystal ErNi$_2$B$_2$C along the c-axis and in the ab-plane. Our results show for the first time the presence of two dominating OPs in ErNi$_2$B$_2$C, which differ by a factor of about two, and appreciable depression of both OPs by the antiferromagnetic transition is found.

\section{Experimental details}

We have used single crystals of ErNi$_{2}$B$_{2}$C grown by the Ames Laboratory Ni$_2$B high-temperature flux
growth method \cite{Cho95}. PCs were established both along the c axis and in perpendicular direction by
standard ``needle-anvil'' methods \cite{Naid}. The ErNi$_{2}$B$_{2}$C surface was prepared by chemical etching
or cleavage as described in \cite{Bobr06}.  As a counter electrode, edged thin Ag wires ($\oslash$=0.15\,mm)
were used to improve mechanical stability of PCs in comparison to use of a bulk Ag piece. We have measured the
temperature dependence of $dV/dI(V)$ characteristics of such N-S PCs (here N denotes a normal metal and S is the
superconductor under study) in the range between 1.45\,K and T$_{c}$ for several contacts oriented both along
the c-axis and in the ab-plane. In the paper we demonstrate results of analysis of 60 $dV/dI(V)$ along the
ab-plane measured for the same PC at different temperatures between 1.45\,K and 11\,K and of 46 $dV/dI(V)$ along
the c-direction for another PC \footnote{The PC resistance is 36 $\Omega$ along the c-axis and 10.5 $\Omega$ in the ab-plane. The PC diameter estimated by Wexler formula (see\cite{Naid}, pages 9, 31) is about 7\,nm and 14\,nm, respectively, using $\rho l\simeq 10^{-11}\Omega$ cm$^2$ \cite{Shulga0}. At the same time a mean free path $l$ is 28\,nm, using  $\rho \simeq 3.5^{-6}\Omega$ cm \cite{Cho95} just above T$_c$. Therefore mentioned PCs are close to the ballistic limit $d<l$.} in the same temperature range.

\section{Results and discussion}

\begin{figure}
\begin{center}
\includegraphics[width=8cm,angle=0]{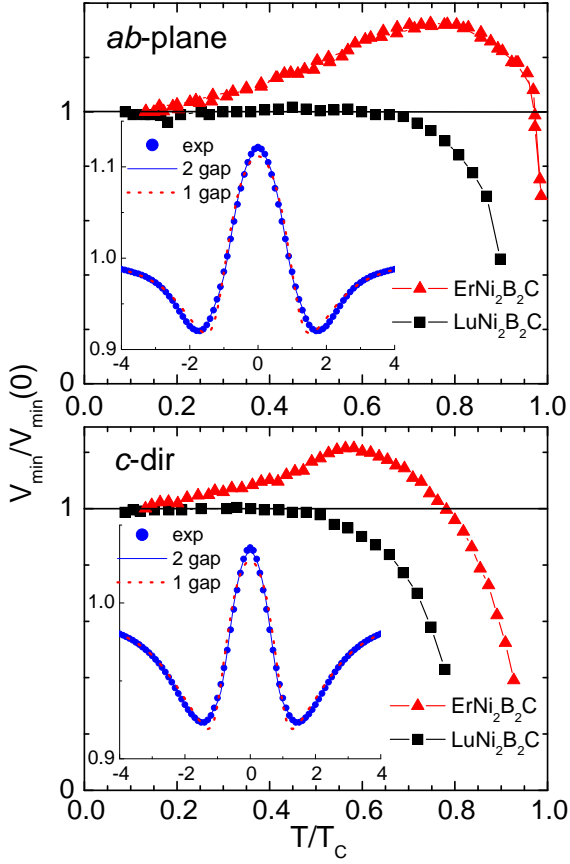}
\end{center}
\vspace{-0.5cm} \caption{(Color online) Reduced position of the minima in the raw $dV/dI$ curves for ErNi$_{2}$B$_{2}$C and that for LuNi$_{2}$B$_{2}$C from \cite{Bobr06}. Insets: comparison of two OPs (solid line) and one-OP (dashed
line) fitting of the reduced experimental $dV/dI$ (symbols) at 1.45\,K.} \label{erf2}
\end{figure}

To determine the OP from the measured differential resistance curves $dV/dI(V)$ we used recent theory \cite{Belob} of Andreev reflection in PC, which includes the pair-breaking effect by magnetic impurities. The last assumption is reasonable, because of the presence of the local magnetic moments of Er ions.
The fit of the measured curves using equations as (1) in \cite{Bobr05} has been performed. As fit parameters the
superconducting OP $\Delta $ \footnote{Assuming that pair breaking is by magnetic impurities, the energy gap
$\Delta_0$ and the OP $\Delta$ are related as follows: $ \Delta_0=\Delta(1-\gamma^{2/3})^{3/2}$ \cite{Belob}.},
the pair-breaking  parameter $\gamma=1/(\tau_s\Delta)$ (here $\tau_s$ is the spin-flip scattering time) and the
dimensionless barrier parameter $Z$ have been used. Although the $dV/dI$ curves shown in Fig.\,\ref{erf1}
exhibit one minimum for each polarity, as in the case of ordinary one gap superconductors \cite{Naid,Naidyuk},
to fit $dV/dI$ in full we had to use a two-OP(gap) approach \footnote{Not only does the one OP
approach give a worse fit (especially at minima position and at maximum, see insets in Fig.\,\ref{erf2}) of the experimental data such that the rms deviation is 2-3 times higher compared to the two OP fit, it also requires a varying $Z$ parameter. On the contrary,
at two OP fit $Z$ parameters remain constant, equal to 0.77 and 0.6 for the ab-plane and c-direction,
respectively. This is important because there is no physical reason for the barrier parameter $Z$ to be
temperature dependent.}, adding the corresponding conductivities as made in \cite{Bobr06} for LuNi$_{2}$B$_{2}$C.
The contributions of these conductivities account for the part of the Fermi surface containing a particular OP.
Thus, for the two-OP model, experimental curves are fitted \footnote{Before the fit, $dV/dI$ curves were
normalized to the $dV/dI$ curve measured above $T_c$ and symmetrized. The fit was done between $\pm$
8\,mV, to avoid contribution from the phonons seen, e.\,g., as an inflection point around 10\,mV for some curves
in Fig.\,\ref{erf1}. The fit was done in two stages. At first we keep both $\gamma_{1,2}$ coefficients equal to
zero. As a result of such fit $\Delta_{1,2}$ values shown in Fig.\,3 were obtained, while variation of $Z$ and
$K$ was within 10\%.  If we will hold $K$ on this stage strictly constant, it results only in more scatter (noise) for $\Delta_{1,2}$, but their overall behavior remains the same. To improve the fit on the second stage, we used obtained $\Delta_{1,2}$ and variate $\gamma_{1,2}$ and $K$. As a result, obtained $K$ and $\gamma_{1,2}$  are shown in Figs.\,4 and 5. After this the theoretical curves were almost indistinguishable from the experimental ones (see insets in Fig.\,2).} by the following
expression:
\begin{figure}
\begin{center}
\includegraphics[width=8.5cm,angle=0]{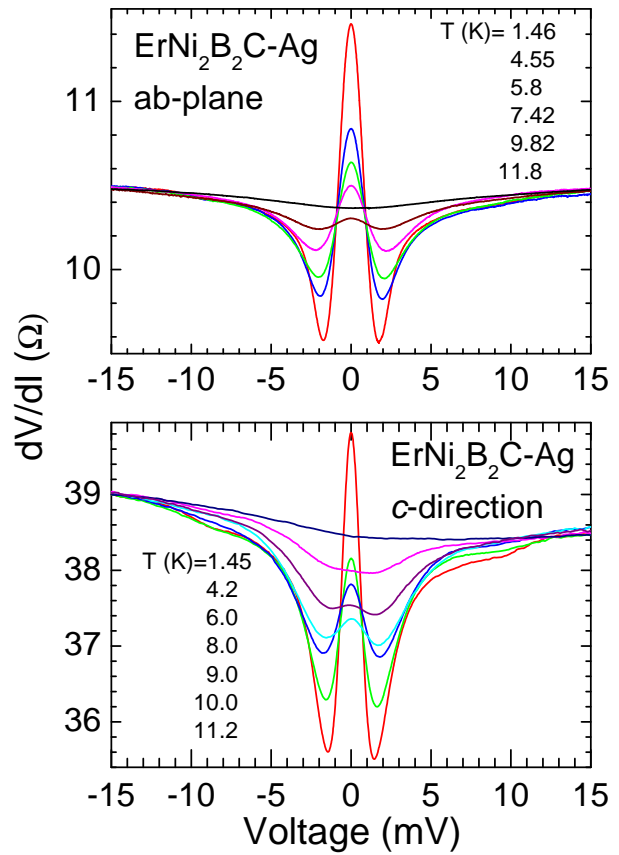}
\end{center}
\vspace{-0.5cm}
\caption{ (Color online) Raw $dV/dI$ curves of ErNi$_{2}$B$_{2}$C--Ag PCs in the ab-plane and
in the c-direction at indicated temperatures. For clarity, only several representative curves
from the total 60 for the ab-plane and 46 for the c-direction measured at different temperatures are shown.} \label{erf1}
\end{figure}

\begin{equation}
\label{twogap}
 \frac{dV}{dI}=\frac{S}{K\frac{dI}{dV}(\Delta_1,\gamma_1,Z_1)+
(1-K)\frac{dI}{dV}(\Delta_2,\gamma_2,Z_2)}
\end{equation}
Here, the coefficient $K$ reflects the contribution of the part of the Fermi surface having the
OP $\Delta_1$, $S$ is the scaling factor to match the amplitude of the calculated and the experimental curves.

Before discussing the fitting results, we point out the unusual specific behavior of the measured $dV/dI$ \footnote{Presented in the paper PCs have survived about 36 hours (c-dir) and 50 hours (ab-plane) of measurements. The $dV/dI$ temperature series for these PCs are the most full, therefore they presented in the paper. Of course, there were other PCs with $dV/dI$ of lower quality or which did not survive temperature sweep in the whole range between 1.45\,K and T$_{c}$. Nevertheless, there were a few of PCs which had $dV/dI$ similar to the presented in the paper, supporting our observations.}
(see Fig.\,\ref{erf1}). First, the distance between $dV/dI$-minima shown in Fig.\,\ref{erf2}, which is often taken as
a rough estimation of the superconducting gap value, increases with temperature before decreasing on approaching
T$_c$ -- quite different behavior from the nonmagnetic LuNi$_2$B$_2$C. Second, the $dV/dI$-minima for
ErNi$_2$B$_2$C persist up to temperatures close to T$_c$ (see Fig\,2, upper panel) though with a small amplitude, leading to a supposition of the presence of the second OP. From these direct observations, a nontrivial behavior of the superconducting OP parameters are expected in ErNi$_{2}$B$_{2}$C.
\begin{figure}[t]
\begin{center}
\includegraphics[width=8cm,angle=0]{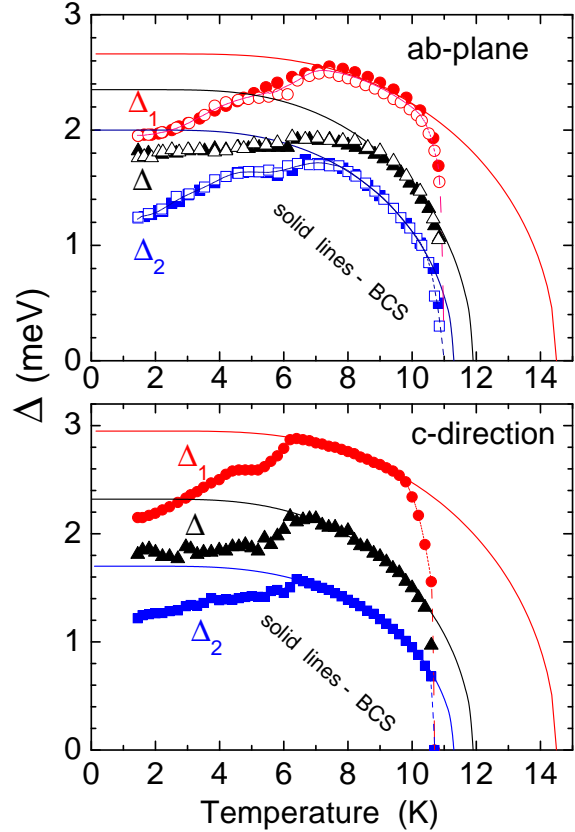}
\end{center}
\vspace{-0.5cm} \caption{ (Color online) Temperature dependence of the large OP $\Delta_1$ (circles), small OP  $\Delta_2$ (squares) and OP $\Delta$ determined by one OP fit (triangles) for ErNi$_{2}$B$_{2}$C for the two main crystallographic directions. In the upper panel closed (open) symbols show OPs determined during
increasing (decreasing) temperature. The same meaning have closed (open) symbols in Figs.\,\ref{erf4} and \ref{erf5}.} \label{erf3}
\end{figure}

Indeed, from the two-band model fitting both OPs $\Delta_1$  (large) and $\Delta_2$ (small) diminish on entering
the antiferromagnetic state around 6\,K (see Fig.\,\ref{erf3}). Qualitatively the same behavior has OP determined by one OP fit (Fig.\,\ref{erf3}, triangles).  This is qualitatively consistent with the
temperature dependence of the superconducting gap determined by tunneling in \cite{Wata}, by laser-photoemission spectroscopy data \cite{Baba} and also with the upper critical field \cite{Budko00,Budko06} and the superconducting coherence length behavior \cite{Gammel99} in the vicinity of T$_{\tiny N}$. The theories of coexistence of superconductivity and antiferromagnetic state also predict such OP suppression below T$_{\tiny N}$,  \footnote{Various theories of antiferromagnetic superconductors including the affect of spin fluctuations, molecular field, and impurities on the $\Delta$ behavior (see e.\ g. H. Chi and A. D. S. Nagi, J. Low Temp. Phys. {\bf 86}, 139 (1992) and Refs. therein) in support of our observation will be discussed in the forthcoming extended publication.} e.\ g., by
antiferromagnetic molecular field \cite{Machida}.

\begin{figure}[t]
\vspace{3cm}
\begin{center}
\includegraphics[width=8.5cm,angle=0]{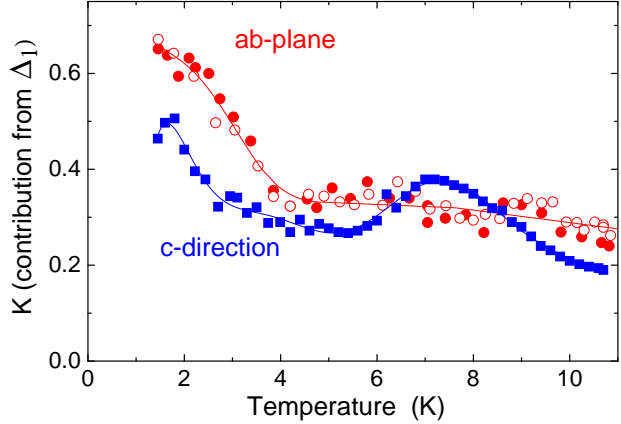}
\end{center}
\vspace{-2.5cm} \caption{(Color online) Temperature dependence of the contribution $K$ (see Eq.\,\ref{twogap})
of the larger OP to the $dV/dI$  spectra. Curves represent a polynomial fit simply to guide the eye.} \label{erf4}
\end{figure}

Further, the large OP $\Delta_1(T)$ may be described by a BCS dependence above T$_{\tiny N}\simeq6$\,K in  the
paramagnetic state with extrapolated T$_{c}^{*}\simeq$14.5\,K, close to that of nonmagnetic $R$Ni$_{2}$B$_{2}$C
($R$=Lu, Y). On the other hand retention of the Andreev reflection minima in $dV/dI$ up to T$_{c}$ (see Fig.\,\ref{erf2}) result in unconventional abrupt $\Delta_1(T)$ vanishing near T$_{c}$.  Note, that to fit experimental curves, not only OP but also large OP relative contribution $K$ (see Eq.\,(\ref{twogap})), must be temperature dependent (see Fig.\,\ref{erf4}). Shown in Fig.\,\ref{erf4}, the decrease of $K$ with temperature points to diminution of
the "superconducting" part of the Fermi surface with the large OP by approaching T$_c$, which results in collapse of the large OP at this point.

\begin{figure}
\begin{center}
\includegraphics[width=8cm,angle=0]{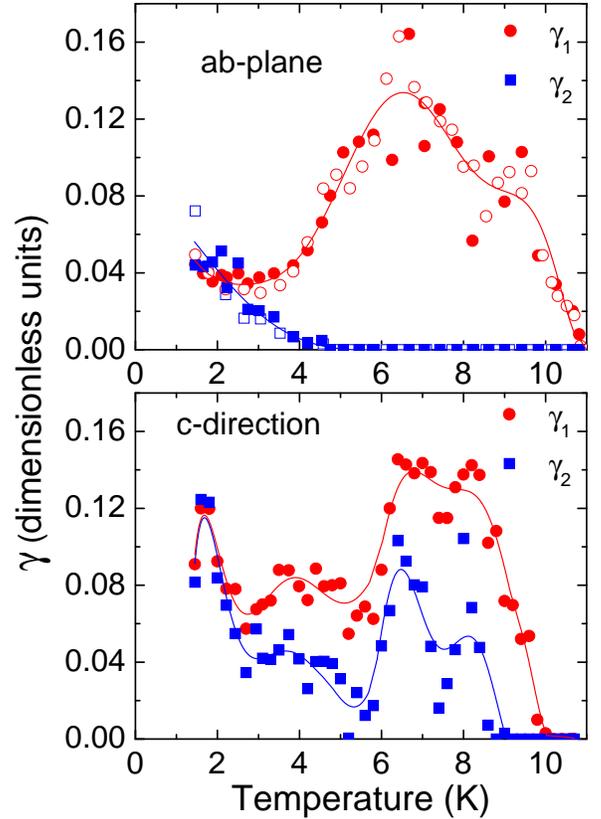}
\end{center}
\vspace{-0.5cm}
\caption{(Color online) Temperature dependence of the pair-breaking parameter $\gamma$. Curves represent polynomial fit simply to guide the eye.}  \label{erf5}
\end{figure}

The mentioned decrease of $K$  correlates with behavior of the pair-breaking parameter $\gamma$ shown in
Fig.\,\ref{erf5}. It appears that $\gamma_1$ is always larger than $\gamma_2$ above 2\,K, that is the
pair-breaking effect is stronger in the band with the larger OP. This is in line with the conclusion, that the
different bands are differently affected by magnetic order, made in \cite{Drechsler,Shorikov} by band structure
analysis of the coexistence of superconductivity and magnetism in the related antiferromagnetic superconductor
DyNi$_{2}$B$_{2}$C, {\em i.e.}, some bands provide a basis for superconductivity while others are important for
the magnetic interactions. Here we should add, that the $S$ parameter in (1) has a maximal value of 0.5 at low
temperature, the same value as the normalized zero-bias tunnel conductivity obtained by STS study of
ErNi$_{2}$B$_{2}$C \cite{Crespo}. So, both observations suggest that nearly half of the total Fermi surface (or
bands) is nonsuperconducting in ErNi$_{2}$B$_{2}$C. Formation of a superzone gap at antiferromagnetic transition
seen in transport measurements \cite{Budko00} may be responsible for this.

From Fig.\,\ref{erf5} it is also seen that $\gamma$ has a maxima at temperature close to $T_{\mbox{\tiny N}}$
and also close to the appearance of weak ferromagnetism around 2\,K, which is reasonable.  At both of these
temperatures an increase in pair breaking is expected due to increasing spin fluctuations accompanying the
corresponding transitions.

\section{Conclusion}
This study demonstrates that the two-band approximation with two OPs, including
pair-breaking effects, suits better for describing the PC Andreev reflection spectra in ErNi$_{2}$B$_{2}$C
%in the whole interval of temperatures below $T_c\simeq$11\,K
pointing for the first time to presence of multiband superconducting OP in this compound. The values and the temperature dependencies of the large and the small OPs have been estimated for the ab-plane and in the c- direction. It is found that in the paramagnetic state both OPs can be described by BCS dependence, but formation of the antiferromagnetic  state below T$_{\tiny N}\simeq6$\,K leads to a decrease of both OPs. The pair-breaking effect is found to be different for the large and the small OP indicating that the different bands are affected differently by magnetic order. This may be the reason for the observed abrupt vanishing of the larger OP at T$_{c}$. It is interesting that extrapolation of the larger OP by ``conventional'' BCS behavior above T$_{\tiny N}$ results in T$_{c}^{*}\simeq$14.5\,K, similar to nonmagnetic YNi$_{2}$B$_{2}$C, so that $2\Delta^{\mbox{\tiny BCS}}(0)/{\rm k}_{\mbox{\tiny B}}$T$_{c}^{*}\simeq4.25$ and 4.7 for the ab-plane and c-direction, respectively. BCS extrapolation gives for the small OP  $2\Delta^{\mbox{\tiny BCS}}(0)/{\rm k}_{\mbox{\tiny B}}$T$_{c}\simeq4.1$ (ab-plane) and 3.5 (c-dir), while for the one OP fit $2\Delta^{\mbox{\tiny BCS}}(0)/{\rm k}_{\mbox{\tiny B}}$T$_{c}\simeq 4.6$ (ab-plane) and 4.5 (c-dir) pointing, in general, to moderately anisotropic and strongly coupled superconducting state in ErNi$_{2}$B$_{2}$C.

\acknowledgments

The support by the State Foundation of Fundamental Research of Ukraine (project $\Phi$16/448-2007),  by the
Robert A. Welch Foundation (Grant No A-0514, Houston, TX), and the National Science Foundation (Grant No.
DMR-0422949) are acknowledged. Ames Laboratory is operated for the U.S. Department of Energy by Iowa State
University under Contract No. W-7405-Eng-82.

\end{document}